\documentclass[fractalfract,article,accept,pdftex,moreauthors]{Definitions/mdpi} 

\firstpage{1} 
\pubvolume{1}
\issuenum{1}
\articlenumber{0}
\pubyear{2024}
\copyrightyear{2024}

\datereceived{ } 
\daterevised{ } 
\dateaccepted{ } 
\datepublished{ } 
\hreflink{https://doi.org/}

\Title{Approaching multifractal complexity in decentralized cryptocurrency trading}

\TitleCitation{Approaching multifractal complexity in decentralized cryptocurrency trading}

\Author{Marcin Wątorek $^{1,3}$\orcidA, Marcin Królczyk $^{1}$\orcidE, Jarosław Kwapień $^{2}$\orcidB,
Tomasz Stanisz $^{2}$\orcidC and Stanisław Drożdż $^{1,2}$\orcidD}

\AuthorNames{Marcin Wątorek, Marcin Królczyk, Jarosław Kwapień, Tomasz Stanisz and Stanisław Drożdż}

\AuthorCitation{Wątorek, M.; Królczyk, M.; Kwapień, J.; Stanisz, T.; Drożdż, S.}

\address{%
$^{1}$ \quad Faculty of Computer Science and Telecommunications, Cracow University of Technology, \mbox{31-155 Krak\'ow, Poland;}\\
$^{2}$ \quad Complex Systems Theory Department, Institute of Nuclear Physics, Polish Academy of Sciences, \mbox{31-342 Krak\'ow, Poland;}\\
$^{3}$ \quad Adapt Centre, School of Computing, Dublin City University, \mbox{D09 Y074 Dublin, Ireland;}}
\corres{Correspondence: stanislaw.drozdz@ifj.edu.pl}

\abstract{Multifractality is a concept that helps compactly grasping the most essential features of the financial dynamics. In its fully developed form, this concept applies to essentially all mature financial markets and even to more liquid cryptocurrencies traded on the centralized exchanges. A new element that adds complexity to cryptocurrency markets is the possibility of decentralized trading. Based on the extracted tick-by-tick transaction data from the Universal Router contract of the Uniswap decentralized exchange, from June 6, 2023, to June 30, 2024, the present study using Multifractal Detrended Fluctuation Analysis (MFDFA) shows that even though liquidity on these new exchanges is still much lower compared to centralized exchanges convincing traces of multifractality are already emerging on this new trading as well. The resulting multifractal spectra are however strongly left-side asymmetric which indicates that this multifractality comes primarily from large fluctuations and small ones are more of the uncorrelated noise type. What is particularly interesting here is the fact that multifractality is more developed for time series representing transaction volumes than rates of return. On the level of these larger events a trace of multifractal cross-correlations between the two characteristics is also observed.}


\begin{document}

\section{Introduction}

There are two basic types of exchanges on the cryptocurrency market: centralized (CEX)~\cite{BonneauJ-2015a} and decentralized (DEX)~\cite{MohanV-2022a,XuJ-2023a}. CEX exchanges operate very similarly to exchanges on traditional financial markets -- their main task is to act as an intermediary between buyers and sellers. The entire exchange infrastructure is completely controlled by a single entity, being often a private company that decides, which elements of the system are publicly available and which should be hidden from ordinary users. The main source of earnings for such exchanges is transaction fees. They are efficient, easy to use, have high liquidity, allow one to exchange fiat money for cryptocurrencies and vice versa, and have well-developed customer service. However, their very nature carries a number of threats: a potential hacker attack on the exchange or dishonesty of its owners may result in a complete loss of funds by its customers~\cite{FederA-2018a,VIDALTOMAS2023,hagele2024centralized}. For this reason, these exchanges often require their users to complete KYC/AML procedures~\cite{MorenoS-2021a}. Currently, the most popular exchanges of this type are Binance, Coinbase Exchange, Kraken, Bybit, Kucoin, OKX, and Bitstamp~\cite{coinmarket}.

In turn, DEX exchanges enable trading without the participation of intermediaries. Their architecture is significantly different from that of centralized exchanges, because the entire infrastructure is controlled by the community. A front-end server is usually open-source code that can be run by anyone. The contract code is fully available for viewing as well. The DEX trading is based on Decentralized Finance (DeFi) -- smart contracts, which are computer programs that are executed without human supervision on a blockchain network like, e.g., Ethereum, which is the most popular one, by using cryptographically signed transactions~\cite{ZhengZ-2020a}. They are concluded only when both parties consent and meet the requirements specified in the contract. Thanks to smart contracts on DEX exchanges, users can trade cryptocurrencies, make loans, and multiply funds through crypto deposits, among others. The advantage of DEX-type exchanges is also the possibility to trade a larger number of tokens, including less popular ones, that are not available on CEX-type exchanges. The decentralized cryptocurrency exchanges are characterized by a high degree of anonymity, because implementing the KYC/AML verification is impossible due to the distributed architecture. Their biggest disadvantages include conscious or unconscious errors, exposure to attacks on smart contracts, the lack of customer support, difficult operation, and often low liquidity and low trading volume on trading pairs~\cite{ASPRIS2021,MohanV-2022a,XuJ-2023a}. The most popular DEX exchanges in 2024 include Uniswap, Radydium, Curve, PancakeSwap, SushiSwap dYdX, Balancer, Orca and Jupiter~\cite{CoinGecko}. In addition to Ethereum, currently popular platforms supporting the operation of smart contracts and DEX are Solana, Tron, BNB Chain, Arbitrum, Base, Avalanche, Sui, Polygon and Optimism~\cite{defillama}.

A key component of the architecture of decentralized cryptocurrency exchanges is the pricing mechanism, based on which three types of DEXs can be distinguished~\cite{ShahK-2023a}:

(1) The Automated Market Makers (AMM), whose task is to determine the appropriate price of assets, are based on smart contracts and liquidity pools (see below)~\cite{BartolettiM-2022a,MohanV-2022a,BroennimannW-2024a}. They are currently the most popular ones.

(2) The exchanges using order books do not have liquidity pools and, in order to determine the price of assets, the traders' orders are matched directly like on the traditional exchanges~\cite{PlattM-2020a}. There are two types of ledgers: the on-chain one that uses blockchain to process data and the off-chain one that uses solutions based on centralization to process transactions~\cite{WarrenW-2017a}. However, both types use blockchain networks to store data.

(3) The DEX aggregators, in which the pricing mechanism involves aggregating asset liquidity from many different protocols~\cite{TakemiyaM-2023a}. Data from many exchanges is concentrated in one place, which allows users to conduct much more profitable trading operations. These exchanges mainly focus on solving the biggest problems of DEXs, which include: low liquidity, inflated cryptocurrency trading prices, long waiting times for the execution of transactions at the price specified by a user, and high prices for service fees.

Liquidity pools are used only in the exchanges based on the AMM protocol~\cite{BartolettiM-2022a,BroennimannW-2024a}. They are a set of funds (DeFi tokens) locked in a smart contract. They are obtained from liquidity providers (i.e., people who deposit their own funds into a pool to enable others to trade) and used by the contract to support decentralized trading. The reward for providing liquidity is receiving remuneration, most often in the form of tokens of a given pool. The remuneration source is most often the transaction fees, which are collected from traders in a given pool.

There are numerous studies that analyzed CEX trading properties~\cite{Corbet2019,MakarovI-2020a,fang2022cryptocurrency}. In particular, the microstructure of the market~\cite{Aleti2021}, the use of machine learning in trading~\cite{Koker2020}, the role of Binance exchange in volatility transmission~\cite{Alexander2022} and the impact of BTC futures contracts on the market~\cite{FASSAS2020} were studied. Exchanges rates on CEX were analyzed from many perspectives~\cite{WatorekM-2021b}, including the noise occurrence~\cite{DIMPFL2021}, optimal trading strategies~\cite{James2022PhysD}, and portfolio construction~\cite{James2023,Nguyen2023}. Multifractal detrended fluctuation analysis (MFDFA) was also applied to study BTC~\cite{TakaishiT-2018a,daSilvaFilho-2018a,StavroyiannisS-2019a,takaishi2020market} and ETH~\cite{MensiW-2019a,han2020long,KakainakaS-2022b,Ali2024} price changes on CEX exchanges. On the other hand, there is a limited amount of such studies for DEX exchanges~\cite{hagele2024centralized}. There are articles that analyze trading properties from the technical point of view~\cite{capponi2021adoption,lo2021uniswap,Alexander2023}, wash trading occurrence~\cite{victor2021detecting,gan2022understanding,gan2024exposing}, arbitrage and maximal extractable value (MEV) possibilities~\cite{WangY-2022a,adams2023don,capponi2024price,hansson2024price}, but there are almost none that analyze the price and volume fluctuations characteristics on DEX, probably due to decentralized character of the data~\cite{hagele2024centralized}. This article covers this unexplored area by systematically analyzing price changes properties on DEX exchanges by advanced methods like MFDFA.

To achieve the research goal the study focuses on transaction data extracted from the Universal Router contract~\cite{UniversalRouter} of the Uniswap DEX~\cite{UniswapDEX}, which is currently the largest DEX exchange in terms of trading volume and number of available tokens~\cite{coinmarket}. It is an automated liquidity protocol based on AMM. Originally, Uniswap operated only on the Ethereum network, but currently, it also operates on Arbitrum, Optimism, Polygon, Base, BNB Chain, Avalanche, Celo, Blast, ZKsync, and Zora~\cite{UniswapChains}. Despite operating on many networks, the exchange is mainly associated with the Ethereum blockchain, and most of its community is concentrated there. Until April 2024, there were four versions of the exchange: Uniswap v1, Uniswap v2~\cite{Uniswapv2}, Uniswap v3~\cite{Uniswapv3}, and the developer version of Uniswap v4 (the full rollout of v4 has been scheduled to take place in the third quarter of 2024)~\cite{Uniswapv4}.

There are two basic types of contracts on Uniswap: (1) Factory contracts -- the so-called base contracts, on top of which other contracts that support the exchange of specific tokens are created. The contract task is also to manage the mapping between the contract address of a given token and the address of its exchange contract. (2) Exchange contracts executing the exchange of a token -- these are contracts that exchange the token assigned to them~\cite{UniversalRouterguide}.

To enable transactions on a given token, it is necessary to perform the following steps: (1) the contract representing a given token must be implemented on the Ethereum network, (2) a trade contract for a given token on the Uniswap exchange has to be created through the Factory contract, and (3) this dedicated contract can be used to execute its exchange transactions. In the 1st version of the exchange, the possibility of creating the token pools was limited as it was possible to create exchange pairs only for the native cryptocurrency Ether (ETH). Currently, however, there are contracts that enable the creation of pools that carry out, for example, a direct exchange of two ERC-20 tokens~\cite{BauerDP-2022a}.

The Universal Router contract~\cite{UniversalRouter} is currently one of the most important contracts on Uniswap. Within a single transaction on the network, this contract can support multiple exchanges of ERC-20 tokens on Uniswap v2 and v3, as well as trading of NFT tokens on multiple markets. The mechanism of combining multiple operations into one transaction offers exchange users great flexibility. It allows, among others, to optimize token exchange costs -- instead of exchanging funds using one low-liquidity pool, a user can, for example, exchange part of their funds on Uniswap v2 and part on Uniswap v3, thus avoiding an unfavorable exchange rate.

The main research question of this article is to investigate to what extent, despite the existing differences in trading methods, the fluctuation characteristics on the most liquid DEX exchange, Uniswap, are similar to those observed on Binance, which is currently the largest CEX exchange. In subsection~\ref{Data}, the characteristics of the log-returns and volume distributions will be compared. The next step will be to check the existence of long-range correlations in the time series of interest rates and volume, which will allow for a comparison of the characteristics of multifractal patterns in time series from both types of exchanges in subsection~\ref{MFDFA_R_V}. Finally, the multifractal properties of correlation between volatility and volume will be investigated in subsection~\ref{MFCCA}.

\section{Data and methodology}

\subsection{Data specification}
\label{Data}

In this study, transactions executed via the Uniswap Universal Router in the most liquid pools are analyzed, specifically: ETH/USDT on Uniswap version 2 (USDT Uv2), ETH/USDT on Uniswap v3 with 0.3\% commission (USDT Uv3\_0.3), ETH/USDT on Uniswap v3 with 0.05\% commission (USDT Uv3\_0.05), ETH/USDC on Uniswap v2 (USDC Uv2), ETH/USDC on Uniswap v3 with 0.3\% commission (USDC Uv3\_0.3), and ETH/USDC on Uniswap v3 with 0.05\% commission (USDC Uv3\_0.05)~\cite{Upools}. The tick-by-tick transaction data covers the period from June 6, 2023, to June 30, 2024. For Uniswap v3 pools, the percentage indicates the fee for the respective pool, whereas for Uniswap v2, the fee is consistently 0.3\%. Uniswap versions 2 and 3 also differ in trading mechanisms. Uniswap v3 introduces concentrated liquidity, allowing liquidity providers to specify the price range within which they provide liquidity~\cite{Uniswapv3}. As a result, each liquidity position may be different and is represented by an NFT token~\cite{UniswapNFT}. Version 3 also includes range orders, enabling orders to be placed within a custom price range above or below the current pool price~\cite{Uniswapv3}. This mechanism can result in transactions with very low volumes and prices that deviate significantly from those on other exchanges. As such, transactions with volumes below 0.01 USD were excluded from the dataset.

\begin{figure}[ht!]
\centering
\includegraphics[width=0.99\textwidth]{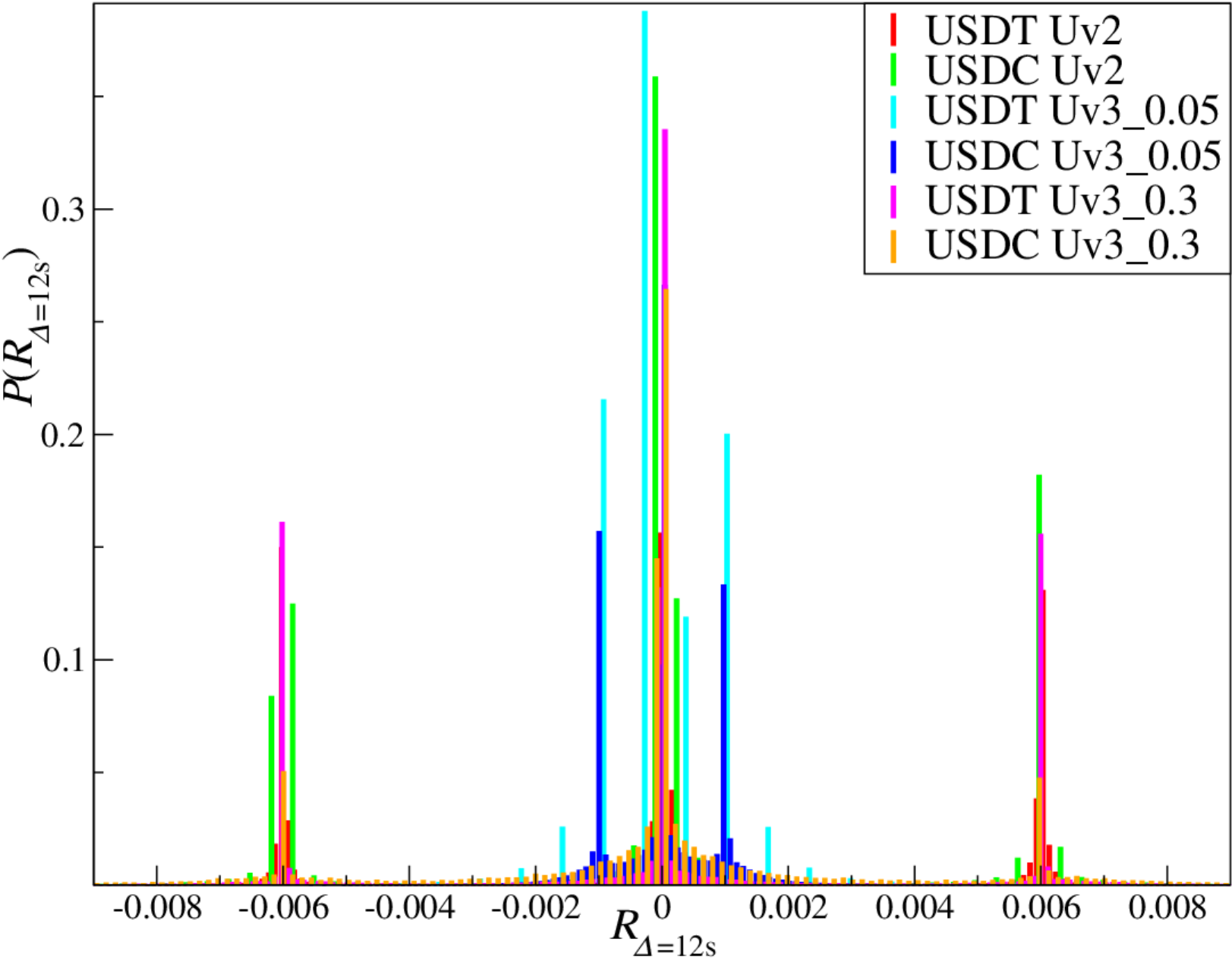}
\caption{The probability distribution (histogram) of the exchange rates ETH/USDT and ETH/USDC log-returns $R_{\Delta t=12s}$ on Uniswap liquidity pools -- versions 2 and 3 with different trading commissions: 0.3\% (USDT Uv3\_0.3, USDC Uv3\_03, USDT Uv2, USDC Uv2) and 0.05\% (USDT Uv3\_0.05 and USDC Uv3\_0.05).}
\label{fig::returns-historgram}
\end{figure}

The distinction between liquidity pools with varying commission rates is evident in the distribution of log-returns, measured at the highest sampling frequency, $\Delta t = 12$s, which corresponds to Ethereum’s block validation time~\cite{Ethereum12s}. Fig.~\ref{fig::returns-historgram} demonstrates that, in addition to the usual concentration of log-returns around zero, common in financial time series~\cite{MantegnaR-1995a,GopikrishnanP-1999a,PlerouV-1999a,Ausloos2000}, notable values of $R_{\Delta t=12s} = 0.006$ and $R_{\Delta t=12s} = 0.001$ appear for pools with 0.3\% (Uv2 and Uv3) and 0.05\% (Uv3) commissions, respectively. This can be attributed to the smallest possible price change (twice the commission rate) at which arbitrage transactions (buying and selling) are profitable~\cite{WangY-2022a}.

To benchmark the time series characteristics from the decentralized exchange Uniswap, exchange rates for ETH/USDT and ETH/USDC from the most liquid centralized exchange, Binance~\cite{Binance}, were also included in the dataset.

The basic characteristics of the time series analyzed in this study are presented in Table~\ref{data-spec}. Notably, the average transaction volume $\langle V \rangle$ is higher on Uniswap than on Binance. Additionally, the maximum observed transaction volume $V_{\textrm{max}}$ is comparable between the two exchanges. However, the average volume over 5min intervals $\langle V_{\Delta t=5\textrm{min}} \rangle$ is significantly higher for the ETH/USDT exchange rate on Binance. This difference can be attributed to the lower number of transactions $N$ on Uniswap compared to Binance. Another important observation is that Uniswap has a significantly lower trading frequency, with the lowest time between transactions $\delta t = 45$s on USDT Uv2, compared to $\delta t = 0.11$s for USDT on Binance.

\begin{table}[]
\caption{Basic statistics of the ETH exchange rate time series considered in this study: total number of transactions $N$, the average inter-transaction time $\delta t$, the average transaction volume value per transaction $\langle V \rangle$, the highest volume value $V_{\textrm{max}}$, the average volume value in a measurement interval $\langle V_{\Delta t} \rangle$, and the fraction of zero log-returns $\%0 R_{\Delta t}$ for $\Delta t$=5min.}
\begin{tabular}{|l|r|r|r|r|r|r|}
\hline
\textbf{Name}       & \textbf{$N$} & \textbf{$\langle \delta t \rangle$ [s]} & \textbf{$\langle V \rangle$ [USD]} & \textbf{$V_{\textrm{max}}$ [USD]} & \textbf{$\langle V_{\Delta t} \rangle$ [USD]} & \textbf{$\%0 R_{\Delta t}$} \\ \hline
USDC Uv3\_0.3  & 37,457          & 900.26      & 9,903                      & 2,358,376                  & 3,298                              & 0.785             \\ \hline
USDC Uv3\_0.05 & 765,399         & 47.01       & 11,227                     & 7,751,943                  & 76,419                             & 0.006             \\ \hline
USDC Uv2       & 503,569         & 66.97       & 1,856                      & 1,254,394                  & 8,310                              & 0.049             \\ \hline
USDT Uv3\_0.3  & 151,003         & 223.33      & 3,390                      & 3,878,300                  & 4,553                              & 0.388             \\ \hline
USDT Uv3\_0.05 & 704,276         & 49.62       & 5,331                      & 2,875,000                  & 33,395                             & 0.009             \\ \hline
USDT Uv2       & 748,919         & 45.14       & 1,843                      & 1,084,474                  & 11,378                             & 0.028             \\ \hline
USDT Binance   & 297,688,280      & 0.11        & 1,185                      & 9,739,140                  & 2,955,093                           & 0             \\ \hline
USDC Binance   & 10,343,176       & 3.31        & 73                        & 4,265,360                  & 72,643                             & 0.003             \\ \hline
\end{tabular}
\label{data-spec}
\end{table}

\begin{figure}[ht!]
\centering
\includegraphics[width=0.49\textwidth]{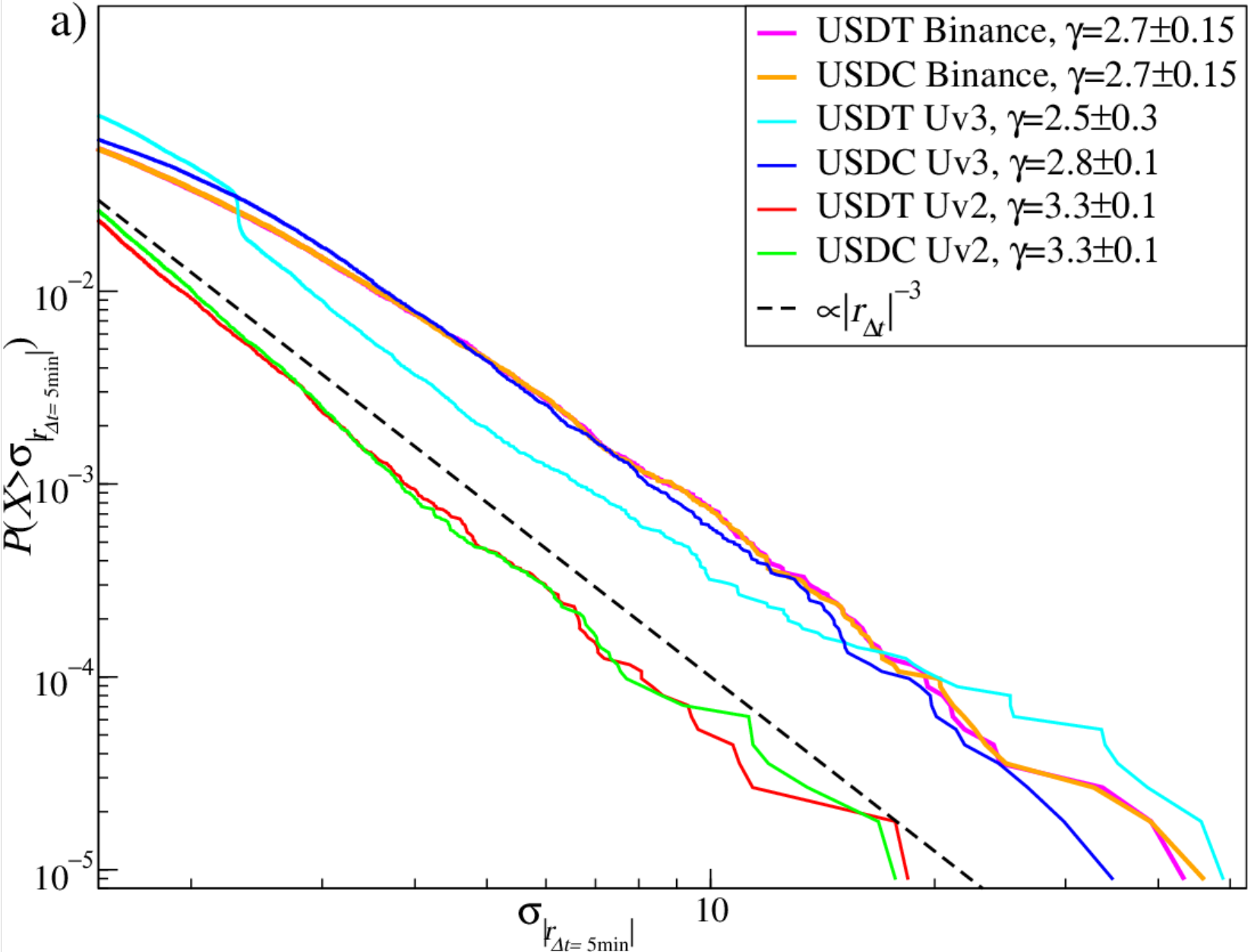}
\includegraphics[width=0.49\textwidth]{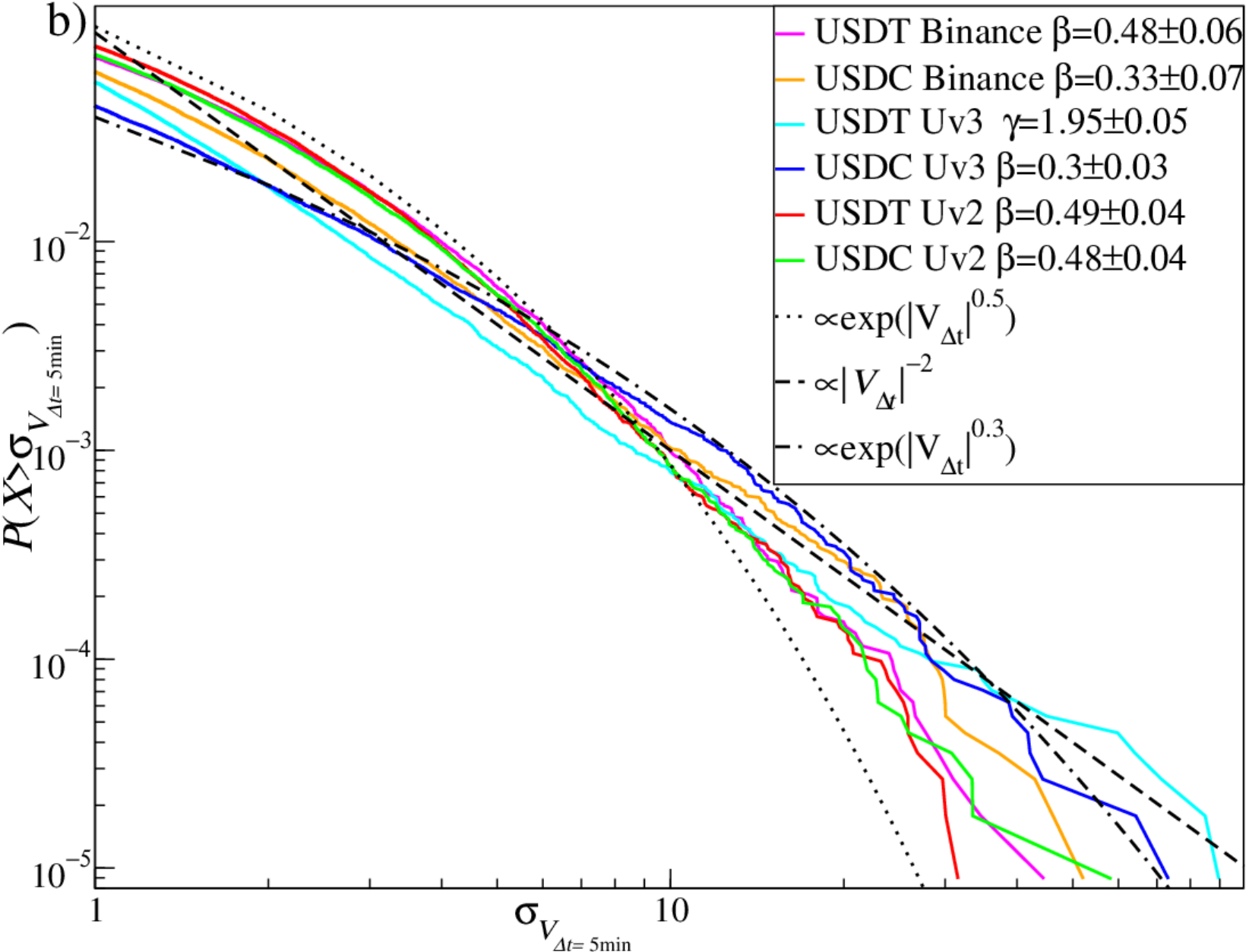}
\caption{Complementary cumulative distribution functions for (a) absolute log-returns $|R_{\Delta \textrm{t=5min}}|$ and (b) volume  $V_{\Delta \textrm{t=5min}}$ of ETH expressed in USDT and USDC on Binance and Uniswap. The estimated exponent, $\gamma$ with standard error, is shown in the insets.}
\label{fig::returns-distr}
\end{figure}

To ensure the analysis remains comparable and to avoid an excessive number of zero returns that could distort more advanced multifractal analysis, the tick-by-tick data was aggregated into a 5-minute time series. Additionally, due to the significantly longer inter-transaction time for Uv3 with a 0.3\% commission ($\delta t > 200s$), the time series from these liquidity pools were combined with those from Uv3 with a 0.05\% commission. Following this procedure, 6 time series of log-returns $R_{\Delta t=\textrm{5min}}$ and volume $V_{\Delta t=\textrm{5min}}$ were obtained for ETH in USDT and USDC on Uniswap v2, Uniswap v3, and Binance: USDT Uv3, USDT Uv2, USDC Uv3, USDC Uv2, USDT Binance, and USDC Binance. Fig.~\ref{fig::returns-distr} presents the complementary cumulative distribution functions (CCDFs) of these time series.

It is a well-established stylized fact that return distributions in financial time series exhibit fat tails~\cite{MandelbrotBB-1963a,Cont2001,Dhesi2021}. This has been documented many times for stock markets~\cite{MantegnaR-1995a,GopikrishnanP-1999a,DrozdzS-2007a}, Forex~\cite{Ausloos2000,GebarowskiR-2019a}, cryptocurrencies~\cite{BEGUSIC2018,James2021b} and even for the most liquid NFT collections~\cite{GHOSH2023,SzydloP-2024a,WatorekM-2024a}. The heavy tails typically manifests as a power-law decay in the complementary cumulative distribution function (CCDF), expressed as $P(X>\sigma) \sim \sigma^{-\gamma}$, where $\sigma$ represents the estimated standard deviation. For high-frequency data, the power-law exponent often approaches a value of 3, i.e., $\gamma \approx 3$~\cite{WatorekM-2021a}. This so-called inverse cubic power law for return distributions has been observed not only in traditional financial markets~\cite{PlerouV-1999a,GopikrishnanP-1999a,DrozdzS-2010a}, but also in the cryptocurrency market~\cite{DrozdzS-2018a,DrozdzS-2019a,DrozdzS-2020a}.

As it is shown in Fig.~\ref{fig::returns-distr}a, this pattern is also evident in the normalized absolute log-returns on the decentralized Uniswap exchange. Interestingly, the tails are fatter, and, thus, the value of $\gamma$ is lower, for Uniswap v3 compared to Uniswap v2. This can be attributed to the more flexible trading mechanism in Uniswap v3, which allows users to set orders within a custom price range above or below the current price. Combined with the ability to provide concentrated liquidity, this results in trades occurring at prices significantly different from those on other exchanges, as observed on March 12, 2024, in the Uv3 ETH/USDT liquidity pool, where the exchange rate reached approximately 4,800~\cite{UniswapblockchianATH}. The $\gamma$ exponent values for Binance exchange rates fall between those observed for the two Uniswap versions.

Another well-known stylized fact that is observed in mature financial markets is that the volume distributions often exhibit fatter tails than the return distributions~\cite{RakR-2013a,YAMANI2023}. In the stock markets, a simple relationship has been proposed between the exponents for volume and log-returns, where $\gamma_V = \gamma_R / 2$~\cite{GabaixX-2003a,PlerouV-2004a}. However, this relationship is more sophisticated for the cryptocurrencies~\cite{Muhammad2020,WatorekM-2023b}, where CCDFs of the normalized volume for the most liquid cryptocurrencies like BTC and ETH are better described by a stretched exponential function $f(x)\sim{\rm exp}(x^\beta)$~\cite{LaherrereJ-1998a}. This finding is supported by the more recent data analyzed in this study. For CCDF of the normalized volume of the most liquid pair on Binance, ETH/USDT, the best fit is $\beta \approx 0.48$ (see Fig.~\ref{fig::returns-distr}b). A slightly fatter tail, and thus a lower $\beta$, is observed for ETH/USDC on Binance.

The observed trend of thicker tails for Uniswap v3 compared to Uniswap v2, seen in the normalized absolute log-returns, also holds for CCDFs of the normalized volume, as shown in Fig.~\ref{fig::returns-distr}b. The heaviest tail, characterized by a power-law decay, is seen in the ETH/USDT pair traded on Uniswap v3, with $\gamma \approx 1.95$. The second heaviest tail is also found for ETH/USDC on Uniswap v3, though its best fit is a stretched exponential function with $\beta \approx 0.3$. In contrast, CCDFs of the normalized volume for ETH/USDT and ETH/USDC traded on Uniswap v2 show thinner tails, with the best fit being $\beta \approx 0.49$.

A more detailed examination of the relationship between price and volume on decentralized exchanges, particularly the concept of price impact~\cite{BouchaudJP-2010a,Wilinski2015}, will be explored in a separate, more specialized article.

\begin{figure}[ht!]
\centering
\includegraphics[width=0.49\textwidth]{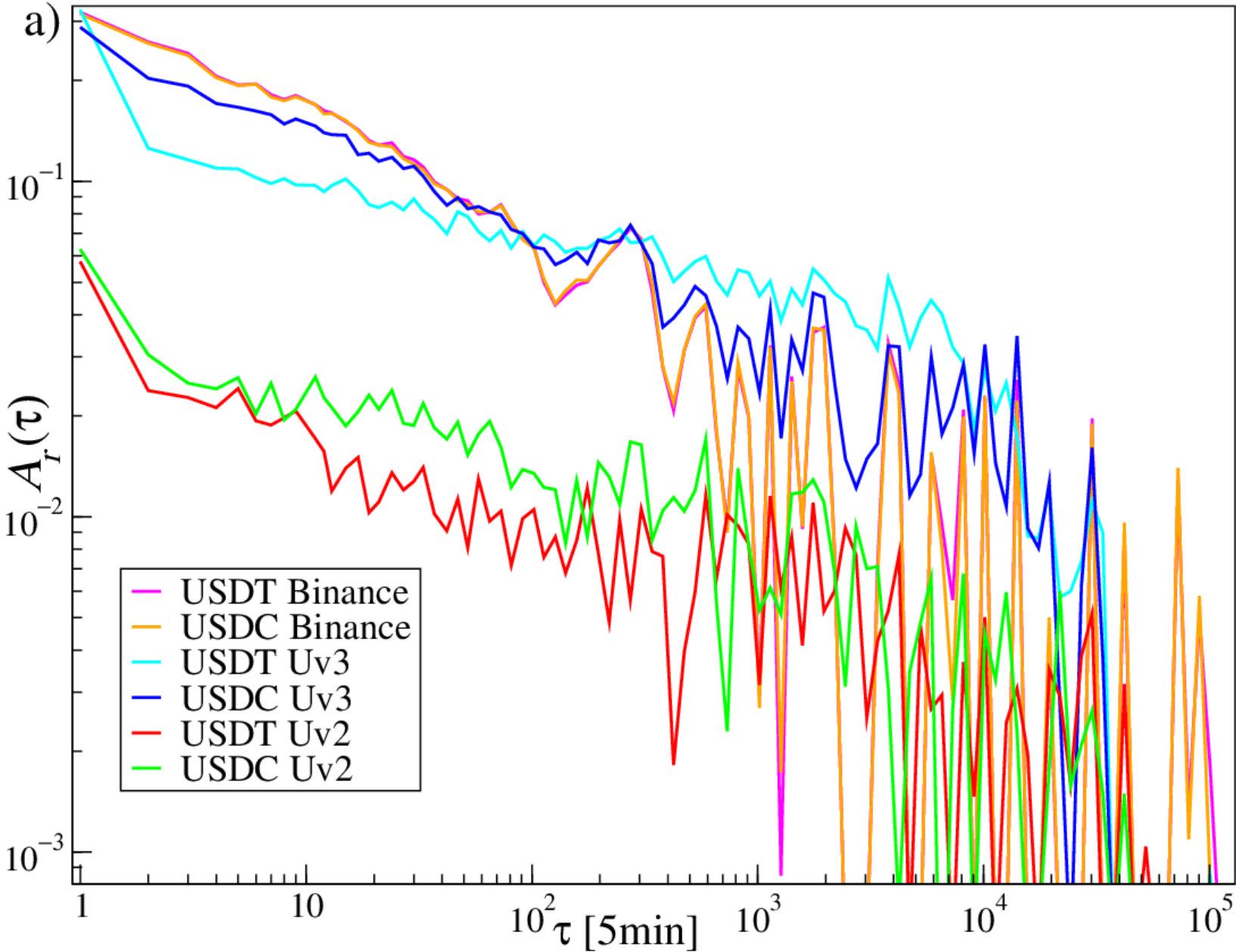}
\includegraphics[width=0.49\textwidth]{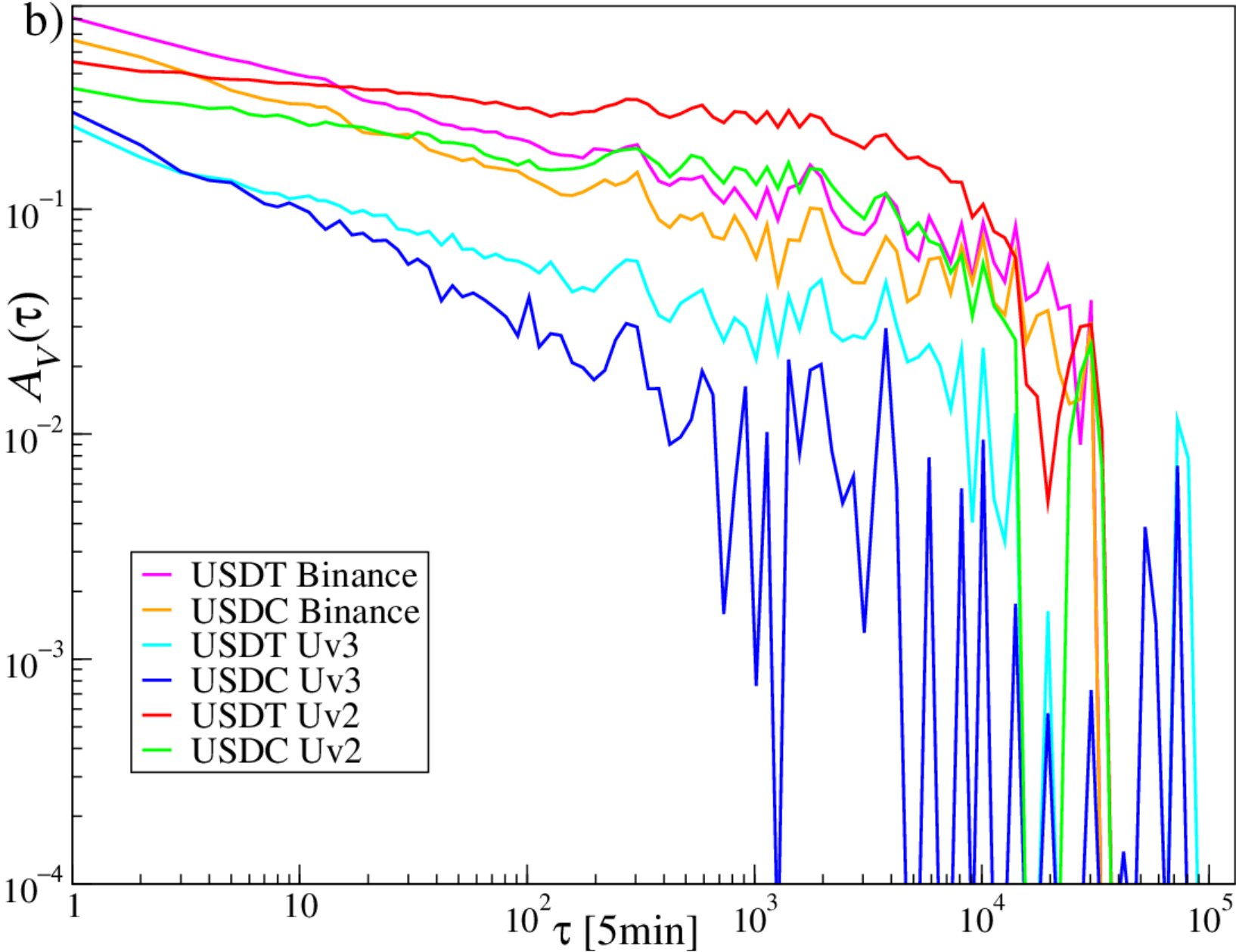}
\caption{Autocorrelation function for a) absolute log-returns $|R_{\Delta \textrm{t=5min}}|$ and b) volume  $V_{\Delta \textrm{t=5min}}$ of ETH expressed in USDT and USDC on Binance and Uniswap exchanges.}
\label{fig::acf}
\end{figure}

The measure used to identify linear correlations in a time series is the autocorrelation function (ACF), defined as:
\begin{equation}
A(x,\Delta i) = { 1/T \sum_{i=1}^T \left[ x(i) - \langle x(i) \rangle_i \right] \left[ x(i+\Delta i) - \langle x(i) \rangle_i \right] \over \sigma^2_x},
\end{equation}
where $\sigma_x$ is the estimated standard deviation of the time series $x(i)$, $\langle \cdot \rangle$ represents the estimated mean, and $\Delta i$ is the time lag in terms of data points, which can be related to the time as $\tau = \Delta i \Delta t$. For financial time series of log-returns, ACF shows a typical pattern: it quickly drops to zero when analyzing the sign of the returns~\cite{Cont2001,Ausloos2000}, but decays slowly, following a power law, when considering the absolute values~\cite{KUTNER2004,RakR-2007a,Klamut2021}. The same behavior is observed in the case of the normalized absolute log-returns on Uniswap, where ACF follows a power-law decay, as it is shown in Fig.~\ref{fig::acf}a.

The strongest autocorrelation, similar to that seen on Binance, occurs in the exchange rates traded on Uniswap v3. In contrast, ACF of the normalized absolute log-returns for Uniswap v2 is significantly weaker, with $A_{r_{\textrm{Uv3}}}(\tau=1) \approx 0.3$ compared to $A_{r_{\textrm{Uv2}}}(\tau=1) \approx 0.06$.

The behavior of ACF for the normalized volume differs from that of log-returns (Fig.~\ref{fig::acf}b). While a power-law decay is also observed, the weakest autocorrelations are found in Uniswap v3, with $A_{V_{\textrm{Uv3}}}(\tau=1) \approx 0.25$. For Uniswap v2, the autocorrelations are slightly stronger, with $A_{V_{\textrm{Uv2}}}(\tau=1) \approx 0.35$ for ETH/USDC and $A_{V_{\textrm{Uv2}}}(\tau=1) \approx 0.45$ for ETH/USDT. Even for larger $\tau$, ACF in Uniswap v2 remains higher than that observed on Binance, where the largest values of ACF occur for $\tau < 20$, with $A_{V}(\tau=1) \approx 0.56$ for ETH/USDC and $A_{V}(\tau=1) \approx 0.7$ for ETH/USDT.

The strength and rate of decay of ACF discussed here influence the scaling of the fluctuation functions, as temporal correlations are a key source of multifractality~\cite{KwapienJ-2023a}. This topic will be explored in detail in the following sections.

\subsection{Multifractal formalism}

Multifractality is a well-documented characteristic observed in financial time series~\cite{Ausloos2002,JiangZQ-2019a,KutnerR-2019a}. It has also been reported in cryptocurrencies traded on centralized exchanges (CEX) such as Binance~\cite{KwapienJ-2022a,KwapienJ-2022b}. The primary objective of this article is to investigate the presence of multifractal effects in the time series of log-returns and trading volume from a decentralized exchange like Uniswap and to examine whether correlations between volatility (expressed by absolute log-returns) and trading volume exist and exhibit multifractality. To test these hypotheses, the multifractal cross-correlation analysis methodology (MFCCA) is applied~\cite{OswiecimkaP-2014a}. This approach is a consistent generalization of the detrended cross-correlation analysis (DCCA) introduced by~\cite{PodobnikB-2008a} and expanded by~\cite{ZhouWX-2008a}. The methodology builds upon multifractal detrended fluctuation analysis~\cite{KantelhardtJ-2002a} (MFDFA), itself an extension of the widely used detrended fluctuation analysis~\cite{PengCK-1994a} (DFA). 

The MFCCA methodology enables a quantitative assessment of both the scaling properties of individual time series and the degree of multifractal cross-correlation between any two time series. The procedure for multifractal cross-correlation analysis involves the following steps. Two time series, $\{x(i)\}_{i=1,...,T}$ and $\{y(i)\}_{i=1,...,T}$, each of length $T$, are divided into $M_s$ segments of length $s$, starting from the opposite ends. The time series are then integrated and, within each segment $\nu$, a least-square-fitted polynomial trend is removed by using the following equations:
\begin{align}
X_{\nu}(s,i) = \sum_{j=1}^i x(\nu s + j) - P^{(m)}_{X,s,\nu}(i), \\
Y_{\nu}(s,i) = \sum_{j=1}^i y(\nu s + j) - P^{(m)}_{Y,s,\nu}(i),
\end{align}
where $P^{(m)}$ represents polynomials of order $m$. The polynomial order of $m=2$ is chosen due to its proven efficacy in financial time series analysis~\cite{OswiecimkaP-2013a}. Following this detrending process, a total of $2M_s$ segments with detrended signals are obtained. Variance and covariance for each segment $\nu$ are then calculated as follows:
\begin{align}
f^2_{\rm ZZ} (s,\nu) = \frac{1}{s}\sum_{i=1}^s (Z_{\nu}(s,i) )^2, \\
f^2_{\rm XY} (s,\nu) = \frac{1}{s}\sum_{i=1}^s X_{\nu}(s,i)\times Y_{\nu}(s,i),
\end{align}
where $Z$ represents either $X$ or $Y$. These quantities are used to compute a family of fluctuation functions:
\begin{align}
F_{\rm ZZ} (q,s) = \biggl\{ {1 \over 2 M_s} \sum_{\nu=0}^{2 M_s-1} \left[ f^2_{\rm ZZ} (s,\nu)\right]^{q/2} \biggr\}^{1/q},
\label{eq::fq.zz} \\
F_{\rm XY} (q,s) = \biggl\{ {1 \over 2 M_s} \sum_{\nu=0}^{2 M_s-1} \textrm{sign} \left[ f^2_{\rm XY}(s,\nu)\right] |f^2_{\rm XY} (s,\nu)|^{q/2} \biggr\}^{1/q},
\label{eq::fq.xy}
\end{align}
where the sign function, $\textrm{sign} \left[ f^2_{\rm XY}(s,\nu)\right]$, ensures consistency of the results across different values of $q$.

Multifractal cross-correlation is expected to reveal itself through a power-law relationship in the scaling behavior of the $q$th-order covariance function $F_{XY}(q,s)$. If it is negative for all $s$, the $-F_{XY}(q,s)$ is taken. The following relationship is obtained:
\begin{equation}
F_{XY}(q,s)\sim s^{\lambda(q)},
\label{Fxyscaling}
\end{equation}
where $\lambda(q)$ is an exponent that quantitatively describes the fractal characteristics of the cross-correlations. In the case of monofractal cross-correlation, $\lambda(q)$ is independent of $q$, while a $q$-dependence of $\lambda(q)$ indicates the presence of multifractal cross-correlations.

$F_{ZZ}$ describes the scaling properties of a single time series through the following relationship:
\begin{equation}
F_{ZZ}(q,s) \sim s^{h(q)},
\label{Fzzscaling}
\end{equation}
where $h(q)$ represents the generalized Hurst exponent~\cite{BarabasiAL-1991a}. If $h(q)$ remains constant, the time series is considered monofractal, with $h(q=2)=H$, where $H$ is the standard Hurst exponent, a measure of the persistence or memory of the time series~\cite{HurstHE-1951a}. When $h(q)$ varies with $q$, the series exhibits a multifractal structure. The corresponding singularity spectrum $f(\alpha)$, which provides further insight into the multifractal nature~\cite{HalseyTC-1986a}, can be computed using the following relations~\cite{KantelhardtJ-2002a}:
\begin{equation}
\alpha = h(q) + q {dh(q) \over dq}, \quad f(\alpha) = q(\alpha - h(q)) + 1,
\label{eq::singularity spectra}
\end{equation}
where $\alpha$ is the singularity (or H\"older) exponent, and $f(\alpha)$ is the multifractal spectrum.
  
This methodology also enables the introduction of the $q$-dependent detrended cross-correlation coefficient $\rho(q,s)$, as proposed by~\cite{KwapienJ-2015a}, which filters out the strength of cross-correlations that vary with fluctuation size:
\begin{equation}
\rho_{\rm XY}(q,s) = {F_{\rm XY}(q,s) \over \sqrt{F_{\rm XX}(q,s) F_{\rm YY}(q,s)}}.
\label{eq::rhoq}
\end{equation}
For $q=2$, this definition can be interpreted as a detrended version of the Pearson cross-correlation coefficient $C$~\cite{ZebendeG-2011a}. In this context, the parameter $q$ acts as a filter: for $q<2$, small fluctuations have more influence on $\rho(q,s)$, while for $q>2$, large fluctuations dominate. This allows the method to selectively emphasize different ranges of fluctuation size contributing to the observed correlations. The filtering capability of $\rho(q,s)$ offers an advantage over the conventional methods since cross-correlations in the real-world time series are typically not uniformly distributed across fluctuations of varying magnitude~\cite{KwapienJ-2017a}. Moreover, $\rho(q,s)$ can be applied to quantify cross-correlations even between the time series that do not exhibit multifractal properties.

\section{Results}
\subsection{Multifractal properties of returns $R$ and volume $V$ time series}
\label{MFDFA_R_V}

\begin{figure}[ht!]
\centering
\includegraphics[width=0.99\textwidth]{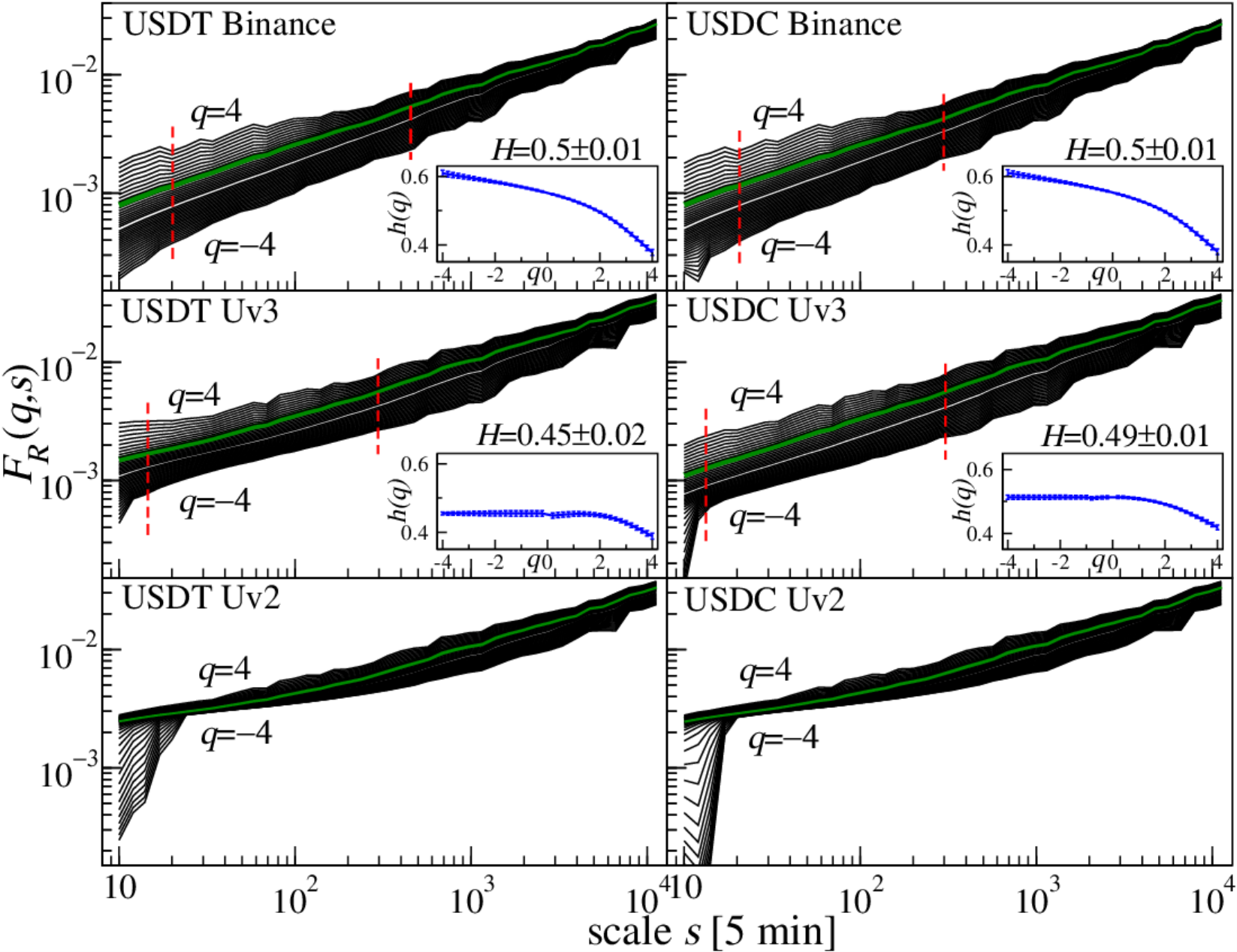}
\caption{Fluctuation functions $F_{RR}(q,s)$ with the range of $q \in [-4,4]$ ($\Delta q = 0.2$) calculated for ETH/USDT and ETH/USDC log-returns $R_{\Delta t=5\textrm{min}}$ from Binance (top), Uniswap v3 (middle), and Uniswap v2 (bottom). (Main) Thick green lines represent $F(q=2,s)$, from the slope of which the Hurst exponent $H$ is estimated together with its standard error. Vertical red dashed lines indicate a scale range, where the family of $F_{RR}(q,s)$ exhibits a power-law dependence for different values of $q$. (Insets) The generalized Hurst exponent $h(q)$ is estimated from the scaling of $F_{RR}(q,s)$. Error bars represent the standard error of linear regression.}
\label{fig::Fq_R}
\end{figure}

\begin{figure}[ht!]
\centering
\includegraphics[width=0.99\textwidth]{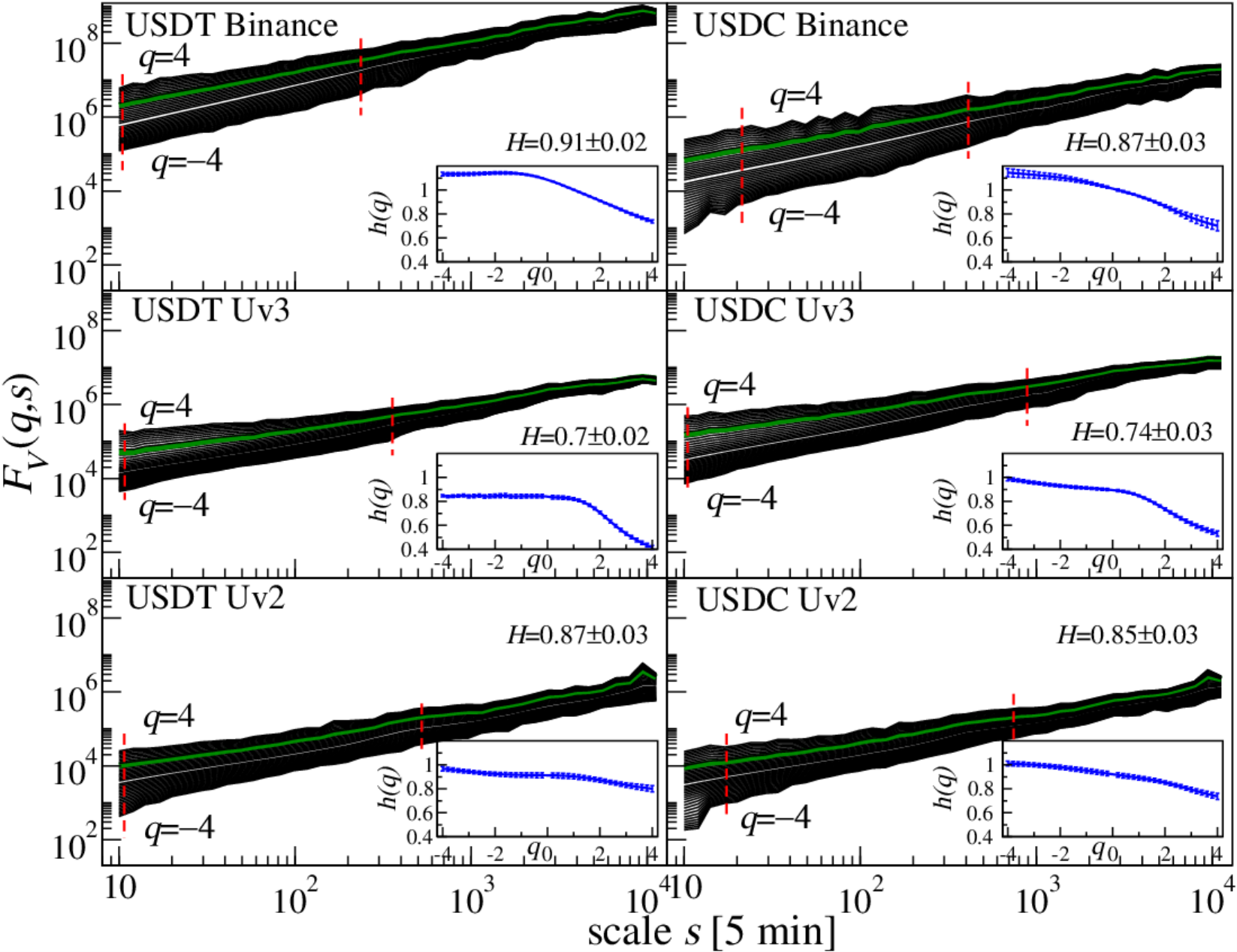}

\caption{$F_{VV}(q,s)$ obtained in the same way as in Fig.~\ref{fig::Fq_R} but for volume $V_{\Delta t=5\textrm{min}}$.}
\label{fig::Fq_V}
\end{figure}

It was shown in Fig.~\ref{fig::acf} that the autocorrelation function for the absolute log-return and volume time series exhibit a power-law decay. The presence of long memory, indicated by this power-law decay, is associated with the occurrence of nonlinear correlations, which are the source of multifractality in time series~\cite{DrozdzS-2009a,ZhouWX-2009a,KwapienJ-2023a}. Therefore, multifractal effects are expected to be present in the analyzed time series. However, the individual ones differ in the strength and persistence of the autocorrelation, which impacts the fluctuation functions $F(q,s)$ calculated using Eq.~(\ref{eq::fq.zz}) and presented in Fig.~\ref{fig::Fq_R} for log-returns and Fig.~\ref{fig::Fq_V} for volume.

For log-returns, the best scaling quality was observed for the exchange rates from Binance (upper panels in Fig.~\ref{fig::Fq_R}). This result is expected, given the behavior of ACF in Fig.~\ref{fig::acf} and the significantly higher liquidity on the Binance pairs compared to the Uniswap ones. Based on the scaling range shown in Fig.~\ref{fig::Fq_R}, the generalized Hurst exponent was estimated using Eq.~(\ref{Fzzscaling}), with the results displayed in the inset of Fig.~\ref{fig::Fq_R} (upper panels). The clear dependence of $h(q)$ across the full range of $q$ indicates that the log-return time series for ETH/USDT and ETH/USDC are multifractal. This finding aligns with previous studies on older data sets for the most liquid BTC and ETH exchange rates~\cite{TakaishiT-2018a,daSilvaFilho-2018a,StavroyiannisS-2019a,MensiW-2019a}. The estimated Hurst exponent values, $H = h(2)$, are close to 0.5, which is typical for mature financial markets~\cite{CARBONE2004,Grech2004,GARCIN2017,MATOS2008,DrozdzS-2018b}.

A slightly weaker scaling of $F_{RR}(q,s)$ is observed for log-returns from Uniswap v3 (middle panels in Fig.~\ref{fig::Fq_R}). In this case, $h(q)$ is dependent on $q$ only for $q > 0$, which corresponds to medium and large fluctuations. Combined with $H<0.5$, this behavior suggests the characteristics typical of immature markets~\cite{DiMatteo2005,Cajueiro2006,James2021,james2022collective,AMMYDRISS2023,Brouty2024}, where multifractality is observed primarily for large fluctuations, while small, noisy fluctuations tend to be monofractal.

For the log-return time series from Uniswap v2 (middle panels in Fig.~\ref{fig::Fq_R}), the scaling quality is so poor that it is not possible to clearly determine the Hurst exponent due to the presence of two distinct scaling regimes in the fluctuation function. This behavior is directly linked to the very weak autocorrelations observed in Fig.~\ref{fig::acf}. A potential reason for the weak autocorrelations and poor scaling of fluctuations in Uniswap v2 log-returns could be the trading mechanism, which does not allow setting orders at arbitrary prices in advance. Instead, trading is only possible by submitting a request to the pool at the current price~\cite{Uniswapv2}.

The scaling quality of the fluctuation functions is better in the case of the volume time series. As it is seen in Fig.\ref{fig::Fq_V}, the scaling of $F_{VV}(q,s)$ is sufficient to determine $h(q)$ for all the time series analyzed. Unlike the log-returns, there is no significant difference between the $F_{VV}(q,s)$ scaling for volume from Binance and both versions of Uniswap. Interestingly, the Hurst exponent for volume is higher for Uniswap v2 ($H \approx 0.86$) compared to Uniswap v3 ($H \approx 0.72$), which is consistent with the stronger autocorrelations observed for Uv2 in Fig.\ref{fig::acf}.

A natural feature of the volume time series is the higher Hurst exponent compared to log-returns, because volume is unsigned. The insets of Fig.~\ref{fig::Fq_V} show that the dependence of $h(q)$ on $q$ is evident for all the volume time series, which indicates their multifractal nature. This dependence is particularly pronounced for $q > 0$, highlighting a more developed multifractality in large fluctuations of the volume time series.

\begin{figure}[ht!]
\centering
\includegraphics[width=0.99\textwidth]{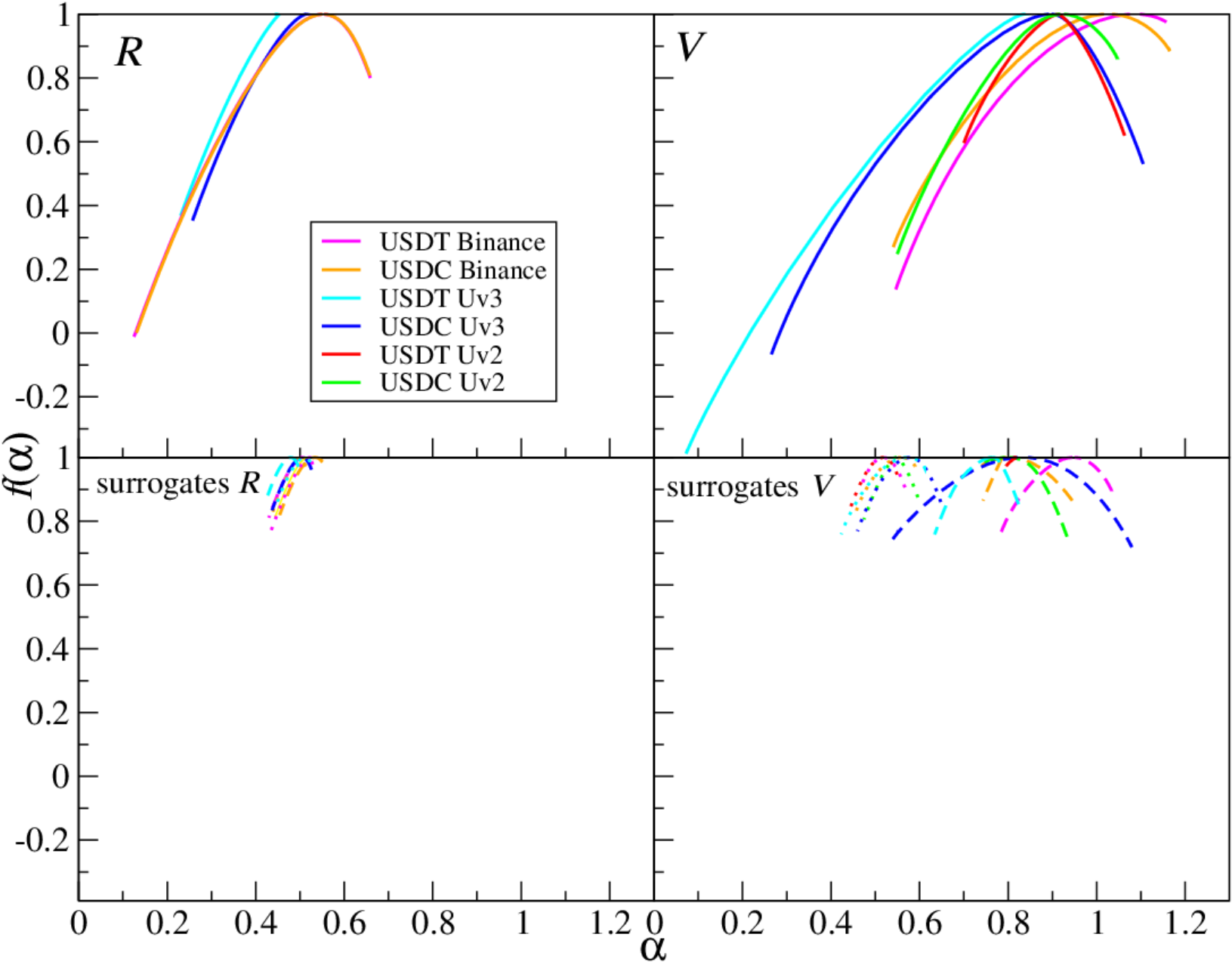}
\caption{Multifractal spectra $f(\alpha)$ calculated for ETH/USDT and ETH/USDC log-returns $R_{\Delta t=5\textrm{min}}$ (left panels) and volume values $V_{\Delta t=5\textrm{min}}$ (right panels) from Binance, Uniswap v3, and Uniswap v2 in the range $q \in [-4:4]$ and $\Delta q = 0.2$. The original time series (top) are compared with their shuffled surrogates marked with dotted lines and the Fourier surrogates marked with dashed lines (bottom).}
\label{fig::spectra}
\end{figure}

This behavior of $h(q)$ is directly reflected in the multifractal spectra shown in Fig.\ref{fig::spectra}, which are derived by using Eq.~(\ref{eq::singularity spectra}). In most cases, a left-sided asymmetry of $f(\alpha)$ is observed, with the left arm stretched relative to the right one. This suggests a richer multifractality in the large fluctuations. Such a left-sided asymmetry in the singularity spectrum is commonly seen in financial time series, where small fluctuations often represent noise, while medium and large ones convey meaningful information~\cite{DrozdzS-2015a,DrozdzS-2018a,JiangZQ-2019a}.

For log-returns (Fig.~\ref{fig::spectra}, left side), the most developed multifractal spectrum is observed for time series from Binance, which confirms previous findings~\cite{WatorekM-2021b,KwapienJ-2022b}. In the case of Uniswap v3 log-returns, only the left arm of the spectrum is present and it is shorter than that of Binance, indicating that only large fluctuations exhibit multifractality. For Uniswap v2 log-returns, the multifractal spectrum could not be determined due to the poor scaling of the fluctuation function $F_{RR}(q,s)$. This suggests that Uniswap, particularly in its version 2, was not yet a mature market from a multifractal perspective. However, in version 3, which was launched later, the appearance of the left arm of the multifractal spectrum may signal a more advanced stage of market development.

Interestingly, for the volume time series (Fig.~\ref{fig::spectra}, right panels), the longest left arm of $f(\alpha)$ is observed for ETH/USDT from Uniswap v3. In the case of ETH/USDC on Uniswap v3, both arms are well-developed. The significant stretching of the left arm for ETH/USDT on Uniswap v3 can be explained by (1) the heavy tails of the corresponding CCDF, where ETH/USDT Uv3 have the heaviest tails (Fig~\ref{fig::returns-distr}b) with $\gamma \approx 1.95$, a value, which causes instability in the Lévy sense, and (2) a slow convergence to the normal distribution~\cite{DrozdzS-2010a}. Additionally, Uniswap v2 also displays multifractality, though $f(\alpha)$ is narrower here than that of ETH/USDC on Uv3. The singularity spectra obtained from the Binance time series show a left-side asymmetry and are comparable to those from Uniswap v2. However, the ones for Binance are shifted to the right indicating higher values of the Hurst exponent, as previously shown in Fig.~\ref{fig::Fq_V}. The smaller differences between Binance and Uniswap in the multifractal characteristics of volume may stem from the fact that, as shown in the basic statistics in Table~\ref{data-spec}, Uniswap has a higher transaction volume per trade, despite its slower trading if compared to Binance.

To assess the statistical significance of the observed multifractal effects in the analyzed time series, two types of surrogate time series were generated: the Fourier surrogates~\cite{SchreiberT-2000a} and the randomly shuffled surrogates~\cite{TheilerJD-1996a}. As it can be seen in the lower panels of Fig.~\ref{fig::spectra}, the multiscale characteristics of the original time series nearly vanish when considering either of the surrogate types. In the case of the Fourier surrogates, nonlinear correlations are eliminated by randomizing the Fourier phases of the original data, leaving only linear correlations intact. As a result, the multifractal spectra $f(\alpha)$ are narrow and shifted to the right for volume since $H>0.5$ for the original time series. For the shuffled surrogates, both the linear and nonlinear correlations are destroyed, though the log-return distributions are preserved. Consequently, $f(\alpha)$ is narrow and centered around $\alpha = 0.5$, which corresponds to an uncorrelated time series ($H=0.5$). It should be noted that the slightly wider spectra for the Uv3 volume are a numerical artifact caused by the heavy tails in the corresponding CCDFs (for more details, see Refs.~\cite{DrozdzS-2010a,GrechD-2013a,ZhouWX-2012a,RakR-2018a}).

\subsection{Correlations between volatility $|R|$ and volume $V$}
\label{MFCCA}

\begin{figure}[ht!]
\centering
\includegraphics[width=0.99\textwidth]{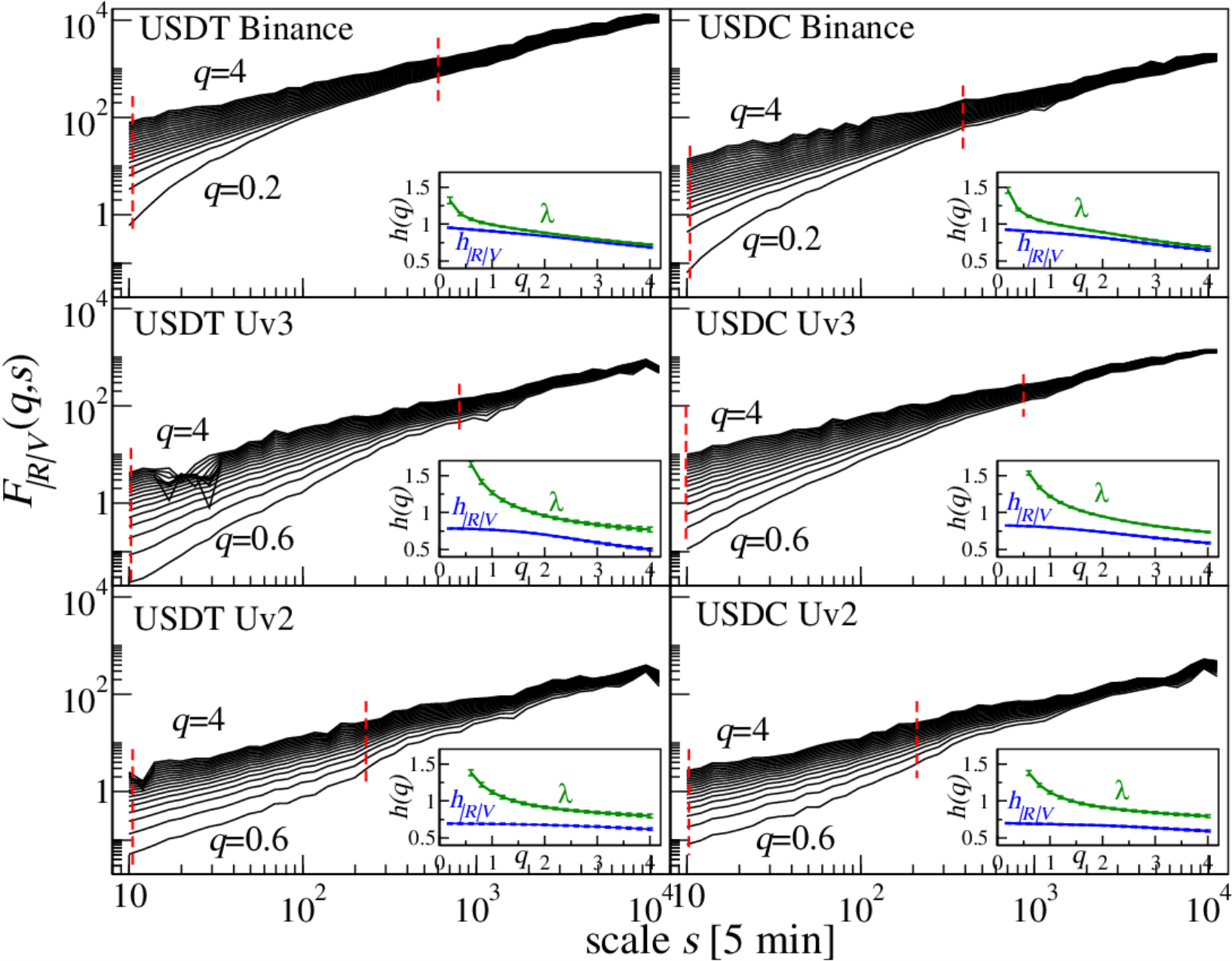}
\caption{(Main) Fluctuation functions $F_{|R|V}(q, s)$ for $q > 0$ with $\Delta q = 0.2$ calculated for volatility $|R_{\Delta 5=\textrm{5min}}|$ and volume $|V_{\Delta 5=\textrm{5min}}|$ representing time series from Binance (top), Uniswap version 3 (Uv3, middle), and version 2 (Uv2, bottom). (Insets) The scaling exponent $\lambda(q)$ and the average generalized Hurst exponent $h_{|R|V}(q)$ are estimated from the range of scales marked with the dashed lines. Error bars represent the standard error of linear regression.}
\label{fig::Fq_RV}
\end{figure}

Another well-documented phenomenon in the mature financial markets is the correlation between volatility (absolute log-returnvolume~\cite{gillemot2006there,podobnik2009cross,RakR-2015a}, which has also been observed recently in the most liquid cryptocurrencies~\cite{WatorekM-2023b}. In this study, the MFCCA methodology will be applied to explore the extent to which volatility and volume are correlated on Uniswap, where the trading mechanisms differ from those on the traditional centralized exchanges. These differences in trading mechanisms have led to the distortions in some of the time series characteristics discussed above.

\begin{figure}[ht!]
\centering
\includegraphics[width=0.99\textwidth]{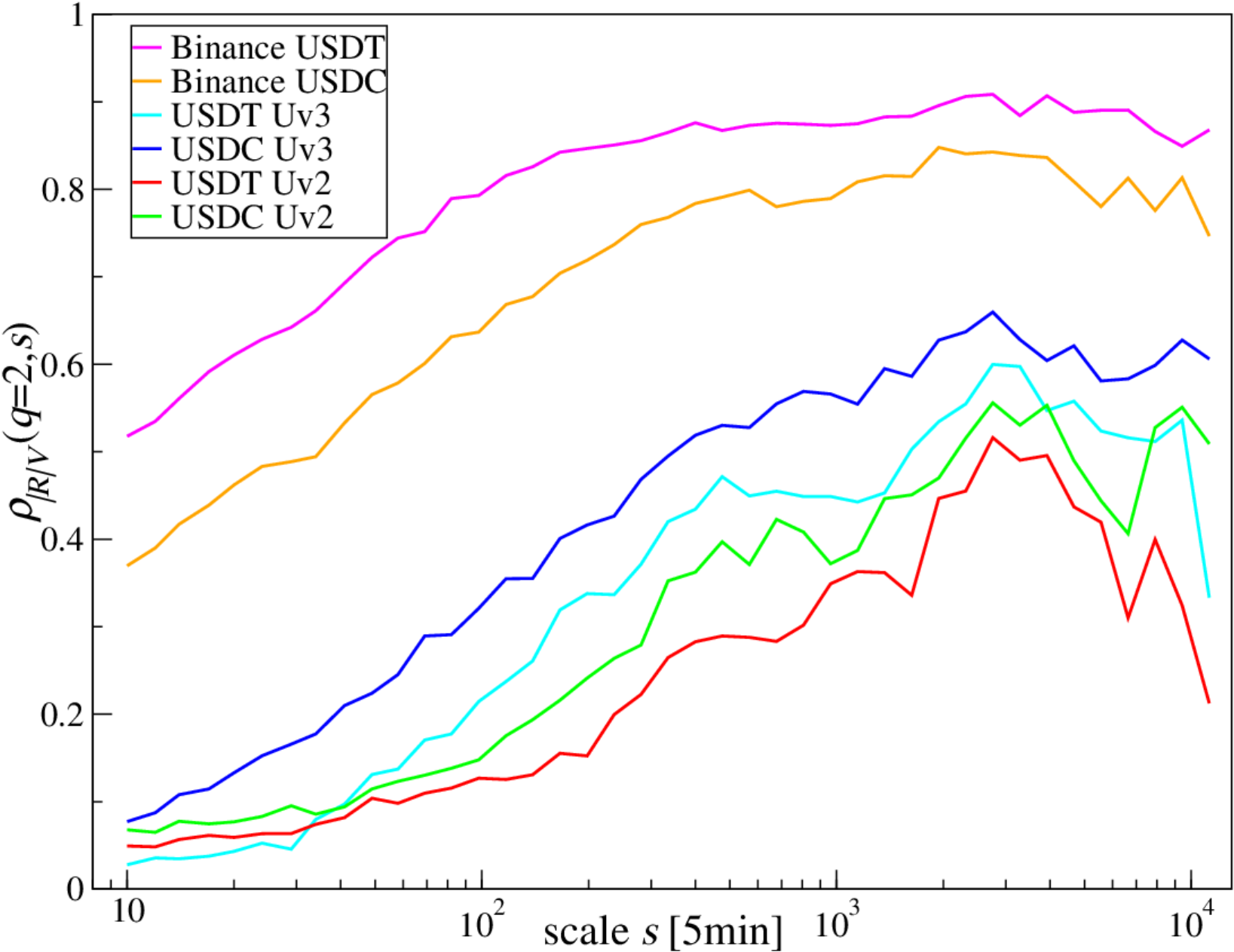}
\caption{The detrended cross-correlation coefficient $\rho(q,s)$, $q=2$, calculated for volatility $|R_{\Delta 5=\textrm{5min}}|$ and volume $|V_{\Delta 5=\textrm{5min}}|$ time series of ETH/USDT and ETH/USDC traded on Binance, Uniswap version 3 (Uv3), and version 2 (Uv2).}
\label{fig::pq_RV}
\end{figure}

The fluctuation functions $F_{|R|V}(q,s)$, representing correlations between volatility $|R|$ and volume $V$, calculated by using Eq.~(\ref{eq::fq.xy}), are shown in Fig.~\ref{fig::Fq_RV}. Similar to the Binance time series (Fig.~\ref{fig::Fq_RV}, top panels), Uniswap also exhibits scaling in the fluctuation functions $F_{|R|V}(q,s)$, but only for $q > 0$, which means that the correlations are present mainly in the medium and large fluctuations. However, the scaling for Uniswap is poorer and begins at higher values of $q$ (around $q=0.6$). This scaling allows one for the estimation of $\lambda$ using Eq.~(\ref{Fxyscaling}) and its comparison with the average generalized Hurst exponents for $|R|$ and $V$ defined as $h_{|R|V}(q) = (h_R(q) + h_V(q)) / 2$~\cite{arianos2009cross} -- see the insets of Fig.~\ref{fig::Fq_RV}. The dependence of $\lambda$ on $q$ is evident, indicating that the correlation between volatility and volume is multifractal. However, in the case of Uniswap, the difference between $\lambda(q)$ and $h_{|R|V}(q)$ is much larger, signaling weaker cross-correlations~\cite{WatorekM-2019a}. This is further reflected in the behavior of the detrended cross-correlation coefficient $\rho(q,s)$ (calculated by using Eq.~(\ref{eq::rhoq})), which is shown for $q=2$ in Fig.~\ref{fig::pq_RV}. For $q=2$ and $s=10$, the coefficient is significantly lower for Uniswap ($\rho(q,s) < 0.1$) compared to Binance ($\rho(q,s) = 0.52$ in the case of ETH/USDT and $\rho(q,s) = 0.37$ in the case of ETH/USDC), particularly at small time scales. In both exchanges, $\rho(q,s)$ increases with time scale $s$, which is in agreement with the related phenomenon observed in the traditional financial markets~\cite{ZhaoL-2018a,WatorekM-2019a,WatorekM-2021b}. However, this increase occurs more rapidly in the time series from Uniswap as $\rho(q,s)$ for Binance plateau around $s=400$, reaching above 0.8.

The strongest correlation in Fig~\ref{fig::pq_RV} is observed for the most liquid ETH/USDT pair traded on Binance, while the weakest correlations are seen in Uniswap v2. This, combined with a poorer scaling of $F_{|R|V}(q,s)$ for Uniswap, highlights the differences between Uniswap and Binance in terms of the volatility-volume cross-correlations. Uniswap appears to be a less developed market in this respect, possibly due to its different trading mechanisms and the fact that transactions on Uniswap occur less frequently but with higher average volumes per transaction than on Binance.

\section{Conclusions}

Despite the different trading mechanisms on the decentralized exchange Uniswap, many of the characteristics of log-returns and volume are already similar to those observed on the centralized exchanges like Binance. However, several differences remain. The log-returns from Uniswap are characterized by weaker autocorrelations, especially for Uniswap version 2, which results in a lack of scaling in the fluctuation function for Uv2 and poor scaling for Uv3. In Uv3, only large fluctuations exhibit a multifractal nature. On the other hand, no significant difference is observed in the volume characteristics between Binance and Uniswap. The cross-correlations between volatility (absolute log-returns) and volume are much weaker on Uniswap. These differences may reflect both the lower maturity level of Uniswap and its unique trading mechanism, where fewer transactions occur but with higher volumes. This newly emerging volatility-volume correspondence on DEX is an intriguing effect from the perspective of the functioning of the financial markets and should definitely become a subject of immediate, more specific, systematic research. The obtained results may help investors to adapt their investment strategies used on more common CEX exchanges to different trading conditions and price-fluctuation properties that occur on DEX exchanges. 

One of the limitations of this study is the fact that the data was obtained only from the Uniswap liquidity pools. Even though it is currently the most liquid DEX exchange, there may exist some specificities on other exchanges. From the investor and arbitrageur perspective, an interesting observation is the significantly higher occurrence of the log returns around the commission-related values. An investigation of this phenomenon on a larger number of liquidity pools and trading protocols will be carried out in the future.

\authorcontributions{Conceptualization, M.W. and S.D.; methodology, M.W., M.K., J.K. and T.S.; software, M.W. and M.K.; validation, M.W., T.S. and S.D.; formal analysis, M.W., M.K., and J.K.; investigation, M.W. and M.K.; resources, M.W. and M.K.; data curation, M.W. and M.K.; writing---original draft preparation, M.W., M.K., J.K., and S.D.; writing---review and editing, M.W., M.K., T.S., and S.D; visualization, M.W. and M.K.; supervision, M.W. and S.D.; project administration, M.W. and S.D.; funding acquisition, M.W. All authors have read and agreed to the published version of the manuscript.}

\funding{This research was funded in part by National Science Centre, Poland, grant number 2023/07/X/ST6/01569.}

\dataavailability{Data is freely available on the public Ethereum blockchain and on Binance platform.} 

\conflictsofinterest{The authors declare no conflicts of interest.} 

\begin{adjustwidth}{-\extralength}{0cm}

\reftitle{References}

\bibliography{refs}

\PublishersNote{}

\end{adjustwidth}

\end{document}